\documentstyle[12pt]{article} 
\include{graphics.sty}
\newcommand{\be}{ \begin{equation}}
\newcommand{\ee}{\end{equation}} 
\begin{document} 
\def\theequation{\arabic{section}.\arabic{equation}} 
\begin{titlepage} 
\setcounter{page}{1} 
\title{Coupled oscillators as models of phantom and scalar field cosmologies} 
\author{Valerio Faraoni\\ \\ 
{\small \it Physics Department, University of Northern 
British Columbia}\\ 
{\small \it 3333 University Way, Prince George, B.C., Canada V2N~4Z9}\\ 
{\small \it  email~~vfaraoni@unbc.ca} }
\date{} \maketitle 
\thispagestyle{empty} 
\vspace*{1truecm} 
\begin{abstract}
We study a toy model for phantom cosmology recently introduced in the literature and 
consisting of two oscillators, one of which carries negative kinetic energy. The results are 
compared with the exact phase space picture obtained for  similar dynamical systems  
describing, respectively, a massive canonical scalar 
field conformally coupled to the spacetime curvature, and a conformally coupled massive 
phantom. Finally, the dynamical system describing exactly a minimally coupled phantom is 
studied and compared with the toy model. 
\end{abstract}
\vspace*{1truecm} 
\end{titlepage}   \clearpage 

\section{Introduction}
\setcounter{equation}{0}

The idea that the universe underwent an inflationary epoch early in 
its history \cite{Guth} has spurred interest in scalar field cosmology. 
A scalar field is  the simplest, although not the unique, way to fuel inflation.
In  this context a scalar field, which may come in different forms and with a variety of 
possible self-interaction potentials, 
is the only or the dominant form of matter in the field equations of the gravitational theory. 
Scalar fields have also been  considered as candidates for dark 
matter and, more recently, for dark energy or quintessence. The recent discovery that the 
universe is undergoing an accelerated expansion that started not long ago \cite{SN} introduces 
the need 
to either modify gravity or postulate a form of dark energy with exotic properties (negative 
pressure). Scenarios based on modified gravity without extra forms of energy 
have thus far not been very  successful. It is easier to describe the observed 
acceleration of the universe by using some form of dark energy, and most models in the 
literature employ a scalar field \cite{darkscalar}.

There is marginal evidence \cite{marginal} that not only the universe accelerates but perhaps 
it even 
superaccelerates, i.e., that the Hubble parameter $H$ may increase with time, $ \dot{H}>0 $, 
which is 
equivalent to an effective equation of state parameter $ w \equiv P/\rho < -1$, where $\rho$ 
and $P $ are the effective energy density and pressure of the dark energy. If 
confirmed, superacceleration  cannot be explained by a canonical,  minimally 
coupled scalar field \cite{FaraoniIJMPD}, but modifications of the theory are required. One of 
such  modifications, considered by many authors \cite{modifiedgravity}, consists in a {\em 
phantom 
field} with 
negative kinetic energy. Such a field is obtained in S-theory, supergravity and in 
higher derivative theories \cite{motiv}. The action for gravity with a phantom 
field $\phi$ is
\be
S^{(phantom)}= \int d^4 x \sqrt{-g} \left[ \frac{R}{2\kappa} +\frac{1}{2} \, g^{ab}\nabla_a 
\phi 
\nabla_b \phi -V( \phi) \right] \;,
\ee
where $\kappa=8\pi G$, $R$ is the Ricci curvature of spacetime \cite{footnote1} and $V(\phi)$ 
is the phantom 
potential. 
Another possibility to explain superacceleration is to let a scalar 
field with positive kinetic energy couple nonminimally to the Ricci curvature of spacetime, as 
described by the action
\be \label{snmc}
S^{(nmc)}= \int d^4 x \sqrt{-g} \left[ \left( \frac{1}{2\kappa} -\frac{\xi}{2}\, \phi^2 
\right) R 
- \frac{1}{2} \, g^{ab}\nabla_a \phi  \nabla_b \phi -V( \phi) \right] \;,
\ee
where $\xi$ is a dimensionless coupling constant. The explicit nonminimal coupling to the 
curvature introduces extra terms in the equations for the scalar field and the cosmological 
metric, allowing for the possibility of superacceleration with  $\dot{H}>0$. A more 
general coupling between the scalar  and gravity, such as the one appearing in 
scalar-tensor and string theories can also generate superacceleration. For example, 
bouncing universes and pole-like inflation, which are not 
possible in general relativity with a minimally coupled canonical scalar, are known in these 
theories \cite{mybook}. Quantum effects can also provide a mechanism to generate 
superacceleration from theories that are classically stable (see, e.g., 
Refs.~\cite{Parkeretal,OnemliWoodard}).

 Apart from the need to explain superacceleration,  
there are other, much more compelling 
reasons to consider a nonminimally coupled field. Nonminimal coupling is introduced by quantum  
corrections to the action of a classical scalar \cite{qc}, and is even required at the 
classical 
level to preserve the Einstein equivalence principle or to avoid causal 
pathologies \cite{SonegoFaraoni}. 

Of particular interest is the case of conformally coupled scalar field theory with $\xi=1/6$. 
In fact, this value of the coupling constant is an infrared fixed point of the renormalization 
group \cite{onesixth} and  therefore we focus on conformally coupled scalars.

Once one accepts the idea of a phantom field with negative kinetic energy, it is natural to 
also consider a nonminimally coupled phantom field \cite{SCK}. While models of nonminimally 
coupled scalars have been studied since the early days of 
inflationary theory, the introduction of phantom fields in cosmology is very recent. 
The  defining feature of a phantom, its negative kinetic energy, has been associated with  
instability and runaway solutions. 
The possibility of negative energies is a well  known feature of scalar-tensor and generalized 
gravity and  seems to be accepted in this context, at least when gravity is classical 
\cite{mybook}. Hence, it is not clear that a negative energy is {\em a priori} a problem. 
Moreover, when a field is coupled to gravity the gravitational energy plays a role in the 
dynamics and no meaningful prescription is available for the 
energy density of the gravitational field even in general relativity. Issues of principle 
with  negative energy are discussed in section~6.  Satisfactory  answers to the questions 
raised, however, come from the understanding of the dynamics, which constitutes the main part 
of this paper.

The problem of the stability of a phantom has 
motivated the consideration of a toy model consisting of two coupled oscillators, one with 
positive and one with negative definite kinetic energy, which exhange energy during motion 
\cite{Carrolletal}. 
The canonical  oscillator with positive kinetic energy is supposed to mimic the gravitational 
field while the second oscillator mimics the phantom field. 
This dynamical system is also an {\em exact} description  of the dynamics of a massive 
conformally coupled scalar field in  a FLRW universe, using 
suitably rescaled variables and conformal time. Section~2 summarizes the equations and 
motivations for the coupled oscillator model, while  section~3  analyzes the dynamics of 
this system. Section~4 considers a similar system of coupled oscillators -- both with 
positive kinetic energy -- which describe exactly  a conformally coupled 
phantom in rescaled variables and conformal time. Having learned a lesson from section~4, we 
can approach in section~5 the exact description of a minimally coupled phantom, without 
resorting to toy models -- the phase space is analyzed in rescaled variables also for this 
case. Section~6 contains a discussion and the conclusions.

\section{Coupled oscillators as cosmological models}
\setcounter{equation}{0}

In this section we summarize the equations of phantom and scalar field cosmology and we show 
how they can be  reduced to a model with two coupled oscillators. We set our analysis in the 
context of a spatially flat Friedmann-Robertson-Walker universe described by the line element 
\be  \label{metric}
ds^2=- dt^2 + a^2(t) \left( dx^2 +dy^2 +dz^2 \right) 
\ee
in comoving coordinates $\left( t,x,y,z \right)$. In this homogeneous and isotropic universe 
the scalar field $ 
\phi (t) $ only  depends on the comoving time and self-interacts through the potential $ V( 
\phi) $. For simplicity it is  assumed that the scalar field is the only form of matter 
present.

\subsection{Phantom cosmologies}

A minimally coupled phantom field in the metric (\ref{metric}) has energy density and 
pressure
\be \label{rhopantom}
\rho_{p} = -\frac{1}{2} \, \dot{\phi}^2 +V( \phi ) \;,
\ee
\be \label{pphantom}
P_p =-\frac{1}{2} \, \dot{\phi}^2 - V( \phi) \;,
\ee
which correspond to the usual expressions for the energy density and pressure of a minimally 
coupled scalar but with the sign of the kinetic energy inverted.  A positive potential $V( 
\phi ) $ is necessary if the energy density $\rho_p$  is to be non-negative. Although this 
requirement  is sometimes dropped \cite{Carrolletal} most relativists are 
reluctant to 
give up the weak energy condition, while  all the other energy conditions are 
progressively being abandoned 
\cite{BarceloVisser}.

 The pressure $P_p$ in eq.~(\ref{pphantom}) is more negative than for a 
canonical scalar field, thus making it possible to achieve a  superquintessence phase with $ 
P_p < -\rho_p  $. A minimally coupled scalar with canonical kinetic energy density $ + 
\dot{\phi}^2/2$ in Einstein gravity cannot achieve $P< -\rho$ \cite{FaraoniIJMPD}. 

The Einstein field equations in the metric (\ref{metric}) and with  the phantom 
$\phi$ as material source are
\be  \label{efe1}
H^2=\frac{\kappa}{6} \left[ -\dot{\phi}^2 +2V( \phi) \right]  \;,
\ee
\be \label{efe2}
\dot{H}+ H^2 =  \frac{\kappa}{3}  \left[ \dot{\phi}^2 + V( \phi) \right]  \;,
\ee
\be \label{efe3}
\ddot{\phi}+ 3H \dot{\phi} - \frac{d V}{d \phi} =0  \;,
\ee
where an overdot denotes differentiation with respect to the comoving 
time $t$. Only two equations in the set  (\ref{efe1})-(\ref{efe3}) are independent -- when 
$\dot{\phi} \neq 0$ one can derive the Klein-Gordon equation (\ref{efe3}) from the other two, 
or from the conservation equation $\dot{\rho}_p+3H\left( P_p +\rho_p \right)=0$ satisfied by 
the phantom.

The field equations (\ref{efe1})-(\ref{efe3}) can be derived from the Lagrangian
\be  \label{L0}
L_0 = 3a\dot{a}^2 +\kappa a^3 \left[ \frac{\dot{\phi}^2}{2} +V( \phi ) 
\right] \;,
\ee
or from the Hamiltonian
\be  \label{H0}
{\cal H}_0  = a^3 \left[  H^2 + \frac{ \kappa}{6}  \, \dot{\phi}^2   
- \frac{ \kappa V( \phi )}{3} \right] \;. 
\ee

A toy model for phantom cosmology consists of two coupled oscillators, one with 
positive-definite and one with negative-definite kinetic energy, described by the 
Lagrangian 
\be  \label{toylagrangian}
L_T =\frac{\dot{x}^2}{2}- \frac{\dot{y}^2}{2}- \frac{m_x^2}{2} \, x^2
-\frac{m_y^2}{2} \, y^2 -
\frac{\mu^2}{2} \, x^2 y^2 \;,
\ee
or by the associated Hamiltonian
\be  \label{toyhamiltonian}
{\cal H}_T  =\frac{\dot{x}^2}{2}- \frac{\dot{y}^2}{2}+ \frac{m_x^2}{2} \, 
x^2
+\frac{m_y^2}{2} \, y^2 +
\frac{\mu^2}{2} \, x^2 y^2 \;.
\ee
A similar model was considered in Ref.~\cite{Carrolletal}. However, these authors choose  the 
opposite sign for the term $m_y^2y^2/2$, and the energy of the phantom turns out to be 
$ \rho_p = - \left( \dot{\phi}^2 + m_{\phi}^2\phi^2 \right) / 2 $ and negative-definite 
instead 
of $ \rho_p = \left( -\dot{\phi}^2 + m_{\phi}^2\phi^2 \right) /2$. While in the toy model the 
negative kinetic and 
potential energy of the ``phantom'' oscillator $y$ can be offset by the positive interaction 
energy $\mu^2 x^2y^2/2$, this is not a possibility for the real phantom field. In the exact 
phantom model, expressions for the interaction energy density between gravitational and 
phantom 
fields and for the gravitational field energy density are not available, and it is desirable 
to keep the energy of the phantom positive and preserve the weak energy condition. This 
explains the difference between our Lagrangian (\ref{toylagrangian}) and that of 
Ref.~\cite{Carrolletal}.

As remarked in Ref.~\cite{Carrolletal}, while the total energy of the system is constant, the 
energy of the phantom oscillator could decrease {\em ad infinitum} and the energy of the other 
oscillator could increase without bound -- the system could not have a stable ground state.
This argument has been presented before in the context of scalar-tensor gravity, which also 
has negative energies \cite{mybook}.

The Euler-Lagrange equations derived from the Lagrangian (\ref{toylagrangian}) are
\be  \label{EL1}
\ddot{x}+\left( m_x^2+\mu^2 y^2 \right) x=0 \;,
\ee
\be  \label{EL2}
\ddot{y} - \left( m_y^2+\mu^2 x^2 \right) y =0 \;.
\ee
The physically interesting case corresponds to a massless graviton and $m_x=0$, while the 
phantom 
field is very light with a mass $m_{\phi} \approx 10^{-33} $eV. 
Eq.~(\ref{EL1}) resembles an harmonic oscillator equation with a real $y$-dependent effective 
mass, while eq.~(\ref{EL2}) exhibits a purely imaginary effective mass $i \sqrt{ m_y^2 
+\mu^2 x^2}$. The first property is 
associated with stability and the second one with instability. By 
changing the sign of the mass term for the phantom, as done in Ref.~\cite{Carrolletal}, one 
obtains an effective mass squared  $ \left( m_y^2 -\mu^2 x^2 \right) $ for the $y$-oscillator, 
which is  positive  and corresponds to stability if $\mu^2 x^2 < m_y^2$. Since  
the phantom mass is constrained to be extremely small, one would have to fine-tune the 
coupling parameter $\mu^2$ to 
achieve stability \cite{Carrolletal}.

In the next section we consider the toy model for a massless phantom obtained by 
setting  $m_x=m_y=0$ and $\mu^2=1$ and described by the dynamical system
\begin{eqnarray}  
\ddot{x} & = & -xy^2  \;,
\label{accix} \\
\ddot{y} & = & x^2y \;, \label{acciy}
\end{eqnarray}
which is invariant under the transformations $\left( x, y \right) \rightarrow
\left( - x, y \right) $ and $ \left( x, y \right) \rightarrow  \left( x,- y \right) $.

\subsection{Conformally coupled scalar field cosmology}

There are many reasons to believe that a scalar field in a curved space couples nonminimally 
to 
the Ricci curvature: nonminimal coupling is induced by quantum corrections to the classical 
action \cite{qc} and is required to renormalize the scalar field theory \cite{CCJ}. In 
particular, conformal coupling is an infrared fixed point of the renormalization group  
\cite{onesixth} and is required by the Einstein equivalence principle and to avoid acausal 
propagation of $\phi$-waves \cite{SonegoFaraoni}. 
Hence, we 
consider a conformally coupled massive scalar field. The equations of motion 
obtained by varying the action (\ref{snmc}) with $\xi=1/6$ and $V=m^2\phi^2 /2$ are 
(\cite{GunzigetalCQG,FosterBlanco}, \cite{newvariables}-\cite{Abramoetal})
\be  \label{reduced1}
\dot{H} + 2H^2 -  \frac{\kappa m^2}{6} \phi^2 = 0 \; ,
\ee
\begin{equation}  \label{reduced2}
\frac{\kappa}{2}\,\dot{\phi}^2  + \kappa H\phi\dot{\phi}
- 3H^2 \left( 1- \frac{  \kappa}{6} \phi^2 \right) +\frac{ \kappa m^2}{2} \,  \phi^2 =0 \, ,
\end{equation}
\be
\ddot{\phi} +3H\dot{\phi} +\frac{R}{6} \phi + m^2 \phi =0 \;, 
\ee
and they incorporate the Hamiltonian constraint
\be
H^2=\frac{\kappa }{3} \, \rho  \; .
\ee
The effective energy density and pressure of the scalar can be identified in three physically 
inequivalent ways \cite{BellucciFaraoniNPB} -- only one of these guarantees conservation 
of energy, and it yields
\begin{equation} \label{density}
\rho=  \frac{\dot{\phi}^2}{2} +\frac{m^2}{2} \, \phi^2 + \frac{1}{2} \,  H \phi    \left( 
H \phi   + 2  \dot{\phi} \right)  \; ,
\end{equation}
\begin{equation}                  \label{pressure}
P=  \frac{\dot{\phi}^2}{2} - \frac{m^2}{2} \, \phi^2 
 -\frac{1}{6}  \left[ 4H   \phi\dot{\phi} + 2 \dot{\phi}^2  + 2 \phi \ddot{\phi} + \left( 
2\dot{H} +3H^2 \right) \phi^2   \right]\; .
\end{equation}       
The expression (\ref{density}) of $ \rho $ is considerably complicated but the problem can be 
reduced to a system of  two oscillators with sharply defined 
energies in a fictitious Minkowski space as  follows. By using the rescaled variables 
\cite{newvariables}-\cite{Castagninoetal}
\be  \label{xy}
x\equiv m a \;, \;\;\;\;\;\;\;\;\;\; y \equiv \sqrt{\frac{\kappa m^2}{6} } \, a  \, \phi
\ee
and the conformal time $\eta $ defined by $ dt = a \, d\eta $, the equations of motion become
\begin{eqnarray}
x'' & = &  xy^2 \;, \label{dyn1} \\
y'' & = &    -x^2 y \;,  \label{dyn2}
\end{eqnarray} 
where a prime denotes differentiation  with respect to $\eta$. Eqs.~(\ref{dyn1}) and 
(\ref{dyn2}) are invariant under the transformations 
$ \left( x, y \right) \rightarrow  \left(- x, y \right) $ and 
$\left( x, y \right) \rightarrow  \left( x,- y \right) $ and they can be derived from the 
Lagrangian 
\be  \label{L1}
L_1 = \frac{(y')^2}{2} -\frac{(x')^2}{2}-\frac{1}{2} \, x^2 y^2 \;.
\ee
The associated Hamiltonian is 
\be
{\cal H}_1=  \frac{(y')^2}{2} -\frac{(x')^2}{2} + \frac{1}{2} x^2 y^2 \;,
\ee
and the Hamiltonian constraint is ${\cal H}_1 =0$. 
Note that it is incorrect to think of $x$ and $y$ as describing gravity and matter, 
respectively, because $y$ is a 
mixture of the gravitational and scalar field degrees of freedom. However, $x=ma$ is 
associated with gravity and it is its effective kinetic energy that is negative here.

Formally, the system (\ref{dyn1}) and (\ref{dyn2}) can be obtained from the system 
(\ref{accix}) and 
(\ref{acciy}) by the exchange of variables 
$\left( x, y \right) \rightarrow  \left( y, x \right) $ and $ t \rightarrow \eta $. Note 
however that the 
variables $x$ and $y$ and the time coordinates $t$ and $\eta$ used in the two cases have 
different 
physical meanings. Moreover, for the toy model (\ref{accix}) and 
(\ref{acciy}) we allow any value of $x$, while in conformal scalar field cosmology the 
restriction $x>0$ applies.


\section{Oscillator model for massless phantom and conformal massive field}
\setcounter{equation}{0}

The system of coupled oscillators (\ref{accix}) and (\ref{acciy}) 
was studied by Castagnino {\em et al.} \cite{Castagninoetal}. Their results are:

\begin{itemize}

\item The fixed points of the system are $ \left( x_0, 0 \right) $ and $ \left( 0, y_0 \right) 
$ 
and they are all unstable. The system of two oscillators does not have stable positions of 
equilibrium.

\item All the orbits in the phase space (apart, of course, from the fixed points) go to 
infinity as $ t\rightarrow +\infty$ with $ y(t) \rightarrow +\infty $ monotonically while 
$ x(t) \rightarrow 0 $ oscillating.

\item There are no cycles and there is no chaos.

\item An asymptotic solution that seems to be generic is
\be
x( t ) \approx \frac{2}{ t } \, \sin \left( \frac{ t^3}{3} \right) \;, \;\;\;\;\;
y(t) \approx t^2 \;.
\ee
The kinetic energy associated with the $x$-oscillator is
\be
K^{(x)}=\frac{(\dot{x})^2}{2} \approx 2 \, t^2 \cos^2 \left( \frac{t^3}{3} \right) 
\ee
and oscillates with divergent amplitude, while the kinetic energy of the $y$-oscillator,
$K^{(y)}= - ( \dot{y})^2 /2 \approx -2 t^2 \rightarrow - \infty $. This corresponds to the  
``instability'' described by Carroll {\em et al.} However, Castagnino {\em et al.} argue on 
the 
basis of their numerical analysis that this solution is an attractor, i.e., it is {\em 
dynamically} stable.

\end{itemize}

Technically, while proving that $x(t) \rightarrow 0$ as $t\rightarrow +\infty$, Castagnino 
{\em et al.} do not prove that $x(t)$ oscillates while going to zero. 
However, this is straightforward to demonstrate. Assume that $ x\rightarrow 0$ monotonically. 
Then, if $ x >0$ the curve representing $x(t)$ has the $t$-axis as horizontal asymptote and 
$\dot{x} < 0$, $\ddot{x} >0$. This contradicts eq.~(\ref{accix}) which guarantees that 
$ \ddot{x}=-xy^2 <0 $. The same reasoning applies if $x < 0$ goes to zero monotonically. 
Hence, $x(t)$ must oscillate.

Let us turn now to the case of a massive conformally coupled scalar field. Castagnino 
{\em et 
al.} actually studied the dynamical system (\ref{accix}) and (\ref{acciy}) but their variables 
are not defined as in our eqs.~(\ref{xy}) -- $x$ and $y$ are exchanged in their definition, so 
their system (\ref{accix}) and (\ref{acciy}) is physically equivalent to our system 
(\ref{dyn1}) and (\ref{dyn2}) and their results are derived using conformal 
time \cite{footnote2}.  The proofs in Ref.~\cite{Castagninoetal} are 
then 
given by considering the limit $\eta \rightarrow 
+\infty$.
 This does not necessarily correspond to large comoving times, which is 
what is instead desired.   For 
example, in a  de Sitter space the scale factor is $a=a_0 \mbox{e}^{Ht}=-\, \frac{1}{H\eta}$ 
and $t\rightarrow +\infty$ corresponds to $\eta \rightarrow 0^{-}$. 
Moreover, if big rip solutions of the kind $a =a_0 \left( t_0-t 
\right)^{-1} $ are present, both comoving and conformal time 
stop at a finite value. However, one can consider 
the least upper bound $\eta_*$ on the possible values of $\eta$ along the orbits of the 
solutions and replace the limit 
$\eta  \rightarrow +\infty$ with 
$\eta  \rightarrow +\eta_*$ in the proofs of Ref.~\cite{Castagninoetal}.

When translated in terms of physical variables $a$ and $\phi$ and in comoving time $t$ and by 
imposing the restriction $x>0$, the interpretation of their results is as follows.

\begin{itemize}

\item The fixed points of the system $ \left( x_0, 0 \right) $  are Minkowski spaces 
and they are all unstable. There are no  attractor points.

\item All the orbits in the phase space (apart, of course, from the fixed points) go to 
infinity as $ t\rightarrow +\infty$ with $ y(t) \rightarrow +\infty $ monotonically while 
$ x(t) \rightarrow 0 $ oscillating. These solutions correspond to universes that expand to 
infinity with scalar field that goes to zero oscillating. This result is consistent with the 
known phenomenology of a nonminimally coupled scalar field: oscillating solutions for $\xi>0$ 
(e.g., \cite{Morikawa}) and non-oscillating ones for $\xi < 0 $ (e.g., \cite{Fukuyamaetal}).

\item There are no cycles and there is no chaos.

\item The asymptotic solution 
\be
x( t ) \approx \frac{2}{ \eta } \, \sin \left( \frac{ \eta^3}{3} \right) \;, \;\;\;\;\;
y(t) \approx \eta^2 
\ee
corresponds to a matter-dominated universe with $a(t) =a_0 t^{2/3}$.

\end{itemize}

An analysis on the lines of that of Ref.~\cite{SCK} leads to the conclusion that the dynamical
system is not integrable for general values of the parameter $\mu$: however, it is not
chaotic. (If a quartic self-interaction for the scalar and a cosmological constant are added
to the scenario, there are special values of the parameters for which the system is integrable
\cite{LakSaha,Rochaetal} -- however, these cases are fine-tuned and not physically relevant.)

\section{Conformally coupled phantom}
\setcounter{equation}{0}

Once one accepts the use of a phantom field, it is natural to consider a nonminimally coupled 
phantom \cite{SCK}. Here we focus on  a massive conformally 
coupled phantom field for which the Klein-Gordon equation is  
\be
\ddot{\phi} +3H\dot{\phi} - m^2 \, \phi -\xi R \phi =0 \;.
\ee

In terms of the rescaled variables $x$ and $y$ of eq.~(\ref{xy}) and of 
conformal time $\eta$, the field equations can be derived from the Lagrangian
\be   \label{l2}
L_2= \frac{(x')^2}{2} +  \frac{(y')^2}{2}  + \frac{1}{2} \, x^2 y^2 
\ee
or from the Hamiltonian
\be
{\cal H}_2= \frac{(x')^2}{2} +  \frac{(y')^2}{2}  - \frac{1}{2} \, x^2 y^2 \;.
\ee
The Lagrangian (\ref{l2}) is actually equivalent (apart from an irrelevant overall sign) to 
the Lagrangian $L_1$ of eq.~(\ref{L1}) in which the conformally coupled scalar is turned 
into a phantom, i.e., in which both gravity and the rescaled phantom have negative kinetic 
energies.

The dynamical system is \cite{SCK}
\begin{eqnarray} 
\label{ddyn1}
x'' & = & xy^2 \;, \\
 \label{ddyn2}
y'' & = & x^2 y \;, 
\end{eqnarray} 
and the energy constraint is ${\cal H}_2=0$. The dynamical system (\ref{ddyn1}) and 
(\ref{ddyn2}) is invariant under any of the 
exchanges 
$ \left( x, y \right) \rightarrow  \left( y, x \right) $, 
$\left( x, y \right) \rightarrow  \left(- x, y \right) $, or
$\left( x, y \right) \rightarrow  \left( x, -y \right) $ 
and the phase space is 3-dimensional in these variables. The 
fixed points are the empty Minkowski spaces $\left( x_0, 0 \right)$. A series of results 
analogous 
to those seen in the previous section hold, and they are proven in the same way.\\\\
{\bf Theorem}~~~~{\em all the (nonstationary) orbits in the phase space go to infinity 
with $x\rightarrow +\infty$.}\\\\
{\bf Theorem}~~~~{\em there are no closed orbits in the phase space.}\\\\
Again, all the 
solutions that are not stationary represent universes expanding  to infinity, all the 
Minkowski fixed points are unstable, and there is no chaos. Of course, the symmetry between 
$x$ and $y$ leads to features that are different from the previous 
case:\\\\
{\bf Theorem}~~~~{\em both variables $x$ and $y$ diverge without oscillations as $\eta 
\rightarrow \eta_{*}$.}\\\\
The proof is as follows.  Consider the function $ I( x, x' ) \equiv x'/x$, which is 
positive along orbits different from the fixed points. By differentiating $ I $ 
along the orbits and using eq.~(\ref{ddyn1}) one obtains
\be
\frac{dI}{d\eta}=\frac{ x^2 y^2 -(x')^2}{x^2} \;.
\ee
The energy constraint ${\cal H}_2=0$ yields
\be
\frac{d I}{d\eta}=  \left( \frac{ y'}{x} \right)^2  \;,
\ee
which is strictly positive along the orbits of non-static solutions.  $ I $ is positive and 
strictly increasing, hence the limit $ 
\lim_{\eta \rightarrow \eta_*} I  $ is either $+\infty$ or $l >0$. 
If the limit is $+\infty$ 
then $ \frac{x'}{x} \rightarrow +\infty $ implies that  $ x' \rightarrow +\infty$ faster 
than $x$, and also $x'' \rightarrow + \infty$. The energy constraint implies that
\be  \label{qqq}
y^2 = \left( \frac{x'}{x} \right)^2
+ \left( \frac{y'}{x} \right)^2 \geq \left( \frac{x'}{x} \right)^2
\ee
diverges. Since $ y^2 \rightarrow +\infty$, $y$ cannot oscillate and  have zeros. Hence we 
have 
one of the two possibilities $ y \rightarrow \pm \infty$. The second derivative
 $y''=x^2 y  $ has the sign of $y$ and is either positive or negative, but it does not change 
sign. Hence $y$ does not oscillate.

If instead the limit of $ I $  is finite and there is a horizontal asymptote, then  
$ I' = \left( y'/x \right)^2 \rightarrow 0^{+} $. 
Since $  I' \geq 0$ and $ I'  \rightarrow 0$, it is $  I''  \rightarrow 
0^{+}$. Now,
\be
\frac{d^2I}{d\eta^2}= \frac{2y'}{x^3} \left( xy''-x'y' \right) = 2 \left[ yy'-\frac{x'}{x} 
\left( \frac{y'}{x} \right)^2\right]  \rightarrow 0^{+} 
\ee
and, by taking the limit of this quantity as $\eta \rightarrow \eta_*$ and using the fact 
that $ x'/x \rightarrow l $ and $ y'/x \rightarrow 0$, one obtains
\be
\lim_{\eta\rightarrow \eta_*} y \, y' =0^{+} \;.
\ee
Now we have $I'' >0$, or $ yy' > I \left( \frac{y'}{x} \right)^2 >0 $. Either $y>0 $ and 
$y'>0$ or $ y< 0 $ in conjunction with $y' <0$. In the first case, in order for $y$ to go to 
zero, $y$ must decrease and it must be $y' < 0$: we then have a contradiction. The same 
reasoning applies to the case $ y<0 $ with $ y' <0$, reaching again a contradiction. Hence the 
limit of $I$ cannot be finite.  This completes the proof.

The system is integrable if $\mu=0$, which describes the trivial case of two free 
non-interacting 
particles. 
In general, however, the system is not Liouville-integrable and in this sense it is 
dynamically complex \cite{SCK}, but it is not chaotic.

Finally, we present the two exact solutions
\be
x_1( \eta) = y_1( \eta) = \frac{ \pm \sqrt{2}}{ \eta-\eta_0} \;,
\ee
\be
x_2( \eta) = - y_2( \eta) = \frac{ \mp \sqrt{2}}{ \eta-\eta_0} \;,
\ee
 where $\eta_0$ is a constant and the sign is chosen so that $x >0$.  The relation $ d t =a \, 
d\eta$ is easily 
integrated to express these solutions in terms of comoving time $t$ and see that they  
represent the expanding or contracting de Sitter spaces 
\be
a(t) = a_0 \,  \mbox{e}^{ \pm \, \frac{m \, t}{\sqrt{2}} } \;, 
\;\;\;\;\;\;\;\;
\phi (t) = \pm \sqrt{\frac{6}{\kappa}} \;. 
\ee
This special value of the scalar $\phi$ corresponds to a divergent effective gravitational 
coupling $\kappa_{eff} \equiv \kappa \left( 1-\kappa \phi^2/6 \right)^{-1} $ and to a 
non-removable curvature singularity in Bianchi~I universes \cite{Abramoetal,VF04}.


\section{Minimally coupled phantom in rescaled variables}
\setcounter{equation}{0}
 
We can now approach {\em exactly} the problem of the minimally coupled massive phantom 
without 
resorting to toy models by using the rescaled variables~(\ref{xy}) and 
compare the equations of motion with those obtained in the previous sections. In terms 
of the variables $x$ and $y$, the equations of 
motion are 
\begin{eqnarray}
 \label{birba1}
x'' & = & -\,\frac{ (x')^2}{x} +3xy^2 \;, \\
\nonumber \\
 \label{birba2}
y'' & = & -\,\frac{ (x')^2 y}{x^2} +3 y^3 +x^2 y \;,  
\end{eqnarray} 
while the energy constraint is
\be   \label{birbaenergy}
(x')^2 + \left( y' - \frac{yx'}{x} \right)^2 =  x^2 y^2 \; ,
\ee
and the (rescaled) effective Lagrangian and Hamiltonian (\ref{L0}) and (\ref{H0}) are
\be
L_0=\frac{1}{x} \left[ (x')^2 +x^2y^2 +\left( y' - \frac{x' y}{x} \right)^2 \right] \;,
\ee
\be
{\cal H}_0=\frac{(x')^2}{x} +   \frac{1}{x} \left( y'-\frac{yx'}{x} \right)^2 -xy^2 \;.
\ee
Eqs.~(\ref{birba1}) and (\ref{birba2}) are rather complicated in comparison to the cases 
studied in section~3 and 4, due to the fact that now the phantom is not conformally coupled.

The fixed points are again Minkowski spaces $\left( x_0, 0 \right)$. In the new variables we 
are able to derive various results.\\\\
\noindent {\bf Theorem}~~~~{\em all the orbits in the 
phase space (apart from the stationary 
points) 
go to infinity.}\\\\
To prove this statement consider the function $F( x,x')= xx'$. By differentiating along the 
orbits of the solutions and using eq.~(\ref{birba1}) one obtains $ F' = 3x^2y^2 > 0 $ on 
all the non-static orbits. Since $F( \eta ) $ is positive and monotonically increasing it is 
either $ \lim_{\eta \rightarrow \eta_*} F( \eta) = C >0 $ or 
$ \lim_{\eta \rightarrow \eta_*} F( \eta) = + \infty $. We show that $F$ cannot tend to a 
finite limit. In fact, if $F( \eta) \rightarrow C$, the asymptotic equation $ xx' \approx C$ 
is satisfied and the asymptotic solution is $x =\sqrt{ 2C\eta +D}$, where $D$ is an 
integration constant. Then 
\be   \label{questa}
 x'' \approx -\frac{C^2}{\left( 2C \eta +D \right)^{3/2}}<0 \;,
\ee
while eq.~(\ref{birba1}) combined with the energy constraint yields
\be
x'' =\frac{1}{x} \left( y'-\frac{yx'}{x} \right)^2  +2xy^2 >0 \;,
\ee
in contradiction with eq.~(\ref{questa}). Hence $ F=xx' \rightarrow + \infty$, 
which implies 
that $x \rightarrow +\infty$. \\\\
\noindent {\bf Theorem}~~~~{\em there are no closed orbits (cycles) in phase space.}\\\\
The proof in Ref.~\cite{Castagninoetal} applies also to this case.
Hence all the solutions (apart from the fixed points) represent universes expanding to 
infinity, all the Minkowski space fixed points are unstable, and there is no chaos. 
Furthermore,\\\\
\noindent {\bf Theorem}~~~~{\em the variable $ y$ diverges without oscillating.}\\\\
To prove this statement, consider $ J (x,x') \equiv x'/x$, which is well defined for $\eta 
\rightarrow \eta_*$ and is positive because $F \equiv xx' >0$ and $x>0$. By 
differentiating along the orbits and using eq.~(\ref{birba1}) one obtains 
\be
J'( \eta) = -2 \left(  \frac{ x'}{x} \right)^2 +3y^2 \;.
\ee
The energy constraint (\ref{birbaenergy}) then yields 
\be
J'( \eta) = \frac{2}{x^2} \left( y' -\frac{yx'}{x} \right)^2 +y^2 >0 \;.
\ee
Hence $ J $ is positive and increasing and the limit of $ J $ is either finite, $ 
\lim_{\eta\rightarrow \eta_* } J =k > 0 $, or plus infinity. The limit cannot be finite, or 
else 
there is a horizontal asymptote of $ J $ and $ J'  \rightarrow 0$, which implies that both 
$ \frac{1}{x^2} \left( y' -\frac{yx'}{x} \right)^2 \rightarrow 0 $ and $y 
\rightarrow 0$. But then the energy constraint (\ref{birbaenergy}) implies that 
\be
 \left( \frac{x'}{x}  \right)^2 =- \frac{1}{x^2} \left( y' - \frac{yx'}{x} \right)^2 + y^2  
\rightarrow 0 \;.
\ee
On the other hand, if $ J \rightarrow k $, it is asymptotically $ x'/x \approx k > 0 $, in 
contradiction with $ x'/x \rightarrow 0$. 

Hence, it must be  $ J =x'/x \rightarrow +\infty$, which implies that $x' \rightarrow +\infty$ 
and, using again eq.~(\ref{birbaenergy}),
\be
y^2 = 
 \left( \frac{x'}{x}  \right)^2 +  \frac{1}{x^2} \left( y' - \frac{yx'}{x} \right)^2 \geq J^2 
 \rightarrow + \infty 
\ee
and $ y \rightarrow \pm \infty$. Moreover, $y$ cannot oscillate or it would go through zeros, 
which is in contradiction with $y^2 \rightarrow +\infty$.

Again, one does not expect the system to be integrable, but there is no chaos.

\section{Discussion and conclusions}
\setcounter{equation}{0}

We are now ready to summarize and discuss our findings and to compare them  with previous 
works.

The dynamical system (\ref{accix}) and (\ref{acciy}) that constitutes a toy  model for a 
minimally coupled massless phantom field and a massless graviton in physical time was studied 
by Castagnino {\em et al.} \cite{Castagninoetal}. Their intent was to study the  
dynamical system describing the physically  different situation of a conformally coupled 
massive scalar in rescaled variables and conformal time. The latter is obtained from the 
system (\ref{accix}) and (\ref{acciy}) by the exchange $\left( x, y \right) \rightarrow 
\left( y, x \right) $ and $ t\rightarrow \eta$.  The variables 
$ x$ and $y$ and the conformal time employed in Ref.~\cite{Castagninoetal} have a 
physical meaning that is quite different from that of the coupled oscillators in the toy 
model for the phantom. 
Carroll {\em et al.} \cite{Carrolletal} considered the toy model for a phantom but added a 
negative mass term to 
the potential in order to stabilize the perturbations of the model -- the system studied 
by Castagnino {\em et al.} corresponds to the special case of that considered by Carroll {\em 
et al.} in which the phantom is massless and the perturbations are always unstable.

The phase space picture of the dynamical system 
(\ref{dyn1}) and (\ref{dyn2}) describing a 
conformally  coupled massive scalar field with canonical kinetic energy density is the  
following \cite{Castagninoetal}. All  the fixed 
points are empty Minkowski spaces and they are 
unstable. There are no cycles and all the orbits of the solutions (except, of course, the 
fixed points) represent universes expanding to infinity, while the scalar field goes to zero 
oscillating. Hence, the matter content (the scalar) tends to infinite dilution in the future 
while the universe expands. This is consistent with the known phenomenology of nonminimally 
coupled scalar fields.

The phase space of the dynamical system (\ref{ddyn1}) and (\ref{ddyn2}) describing 
a conformally  coupled massive phantom field is qualitatively different: there are 
again unstable fixed points representing  Minkowski spaces and all the other orbits go to 
infinity, but the $y$-variable diverges  without oscillating. Again, there are no cycles.

We have studied also the exact phase space of a massive, minimally coupled phantom in the 
rescaled $x$- and $y$-variables and using conformal time, without resorting to toy models. 
In these variables, the Lagrangian and Hamiltonian are considerably more complicated than 
in the  previous cases, due to the fact that the $x$- and $y$- variables are suited for 
conformal coupling. Nevertheless, by using these variables we can compare our results 
for a minimally coupled massive phantom with those for conformally coupled canonical scalars 
and 
phantoms.  The results for this case are as follows. There are again Minkowski space fixed 
points, which  are unstable. All the other orbits go to infinity with $y$ diverging without 
oscillations, and  there are no cycles. Allowing the phantom to be massive does not constrain 
the orbits to a finite region of the phase space. In fact the phantom ``falls up'' in the 
potential $m^2 \phi^2 /2$, which does not have a maximum in which $\phi$ can settle.

There is a difference in the qualitative behaviour of the solutions of the exact 
equations for the phantom in this case and those of  the toy 
model discussed in Refs.~\cite{Carrolletal,Castagninoetal}. While in the exact case 
studied here $x$ and $y$ diverge without oscillating, the toy model has $x \rightarrow 0$ with 
oscillating behaviour while $ y \rightarrow + \infty$. The toy  model is clearly inadequate 
and its solutions are wildly different from the solutions of the equations describing the 
actual physical  system comprising the gravitational field and  the phantom.

Finally, let us comment on the instability associated with negative energy which  motivated 
the introduction of a toy model in the first place \cite{Carrolletal}. 
The  argument that a negative kinetic energy leads to instability is recurrent in 
Brans-Dicke, scalar-tensor and  non-linear gravity theories and many 
authors regard it as a rationale to discard the Jordan  frame formulation of scalar-tensor  
theories in favour of their Einstein frame counterpart (see Ref.~\cite{FGN} for a review and 
Ref.~\cite{mybook} for a discussion). This 
point of view is legitimate in flat space physics but is not compelling when gravity is 
included in the picture. First, one cannot write expressions for the energy density of the 
gravitational field (corresponding to the $x$-oscillator in the toy model) and the interaction 
energy density between the two 
fields (corresponding to the $\mu^2x^2y^2/2$ term), and it is not clear how such a separation 
could be achieved. 
This fact causes ambiguities in the definition of energy density for a scalar that couples 
nonminimally to the curvature through an explicit 
non-minimal coupling term or a Brans-Dicke-like term in the action 
\cite{BellucciFaraoniNPB,mybook}. 
Indeed, one can associate unambiguously an energy to scalar waves in 
linearized Brans-Dicke gravity, and this energy turns out to be negative -- hence one would 
expect an instability. However, the Minkowski background is  stable \cite{Faraonisubmitted}. 
Second, the discussion of runaway solutions associated with a 
negative energy instability  becomes delicate in the cosmological context. A 
runaway solution representing a universe that expands forever or recollapses is 
undistinguishable from an ``ordinary'' cosmological solution with the same properties, which 
is regarded as perfectly acceptable. It is true that negative energies in scalar-tensor 
gravity lead to trouble with the formulation of the Cauchy problem or with the quantization of 
the linearized gravitational field \cite{mybook}, but the negative energy cosmological  
solutions seem  acceptable at the classical level. 

When the phantom is treated as a quantum field,  the instability becomes more worrisome 
because, in general, a quantum field has many more states than a particle and there are many 
more channels for decay -- this plays a role in how an instability manifests itself. A 
quantum phantom field can decay into ordinary particles and other 
phantoms and, vice-versa, particles can decay into phantoms and other particles. Since the 
direct coupling of the phantom to ordinary matter must be suppressed to a high degree in order 
not to violate the equivalence principle, there remains the possibility of decays of phantoms 
into gravitons and vice-versa. Since the classical phantom coupled to the gravitational field 
would seem to produce negative energy scalars and positive energy gravitons, and gravity 
couples to the stress-energy tensor of the phantom, the decay process would be ultra-fast 
without a cutoff. Carroll {\em et al.} \cite{Carrolletal} study this 
possibility and conclude that it is not unreasonable to obtain lifetimes larger than the age 
of the universe if a momentum cutoff as low as $10^{-3}$eV can be motivated. At present, there 
is no justification for such a cutoff, while a natural cutoff would instead be at the Planck 
scale. Hence, although at the classical level the purported instability of the classical 
unperturbed model does not seem to be fatal for phantom cosmology, at the quantum level the 
survival of the universe containing a phantom for 13.7 billion years looks surprising.

\section*{Acknowledgments}

It is a pleasure to thank Dr. Sebastiano Sonego for stimulating discussions and  an anonymous 
referee for suggestions leading to improvements in the manuscript. This work was supported by the 
National Science and Engineering Research Council of Canada ({\em NSERC}) through a  Discovery Grant.


\begin{thebibliography}{99}
\bibitem{footnote1} We use the metric 
signature $-,+,+,+$ and units in which the speed of light is unity. $G$ is Newton's 
constant and the other conventions follow Ref.~\cite{Wald}.
\bibitem{footnote2} Beware of a typo in Ref.~\cite{Castagninoetal}: the authors define $x 
\equiv  m\phi$,  which should instead read $x \equiv m a\phi$.
\end{thebibliography}
\end{document}